\documentstyle[editedvolume,numreferences,epsfig]{jd} 

\begin{opening}
\title{Deep Surveys and Cosmology}
\subtitle{}

\author{S. J. Oliver}
\author{S. Sergeant}
\author{P. Goldschmidt}
\author{R.G. Mann}
\author{M. Rowan-Robinson}
\author{N. Eaton}
\author{A. Efsathiou}
\author{C. Gruppioni}
\author{T.J. Sumner}
\author{B. Mobasher}
\author{A. Verma}
\author{L. Danese}
\author{E. Egami}
\author{D. Elbaz}
\author{A. Franceschini}
\author{I. Gonzalez-Serrano}
\author{M. Kontizas}
\author{A. Lawrence}
\author{R. McMahon}
\author{H.U. N{\O}rgaard-Nielsen}
\author{I. P$\acute{\rm e}$rez-Fournon}

\institute{Astrophysics Group\\
           Imperial College of Science Technology and Medicine}

\end{opening}

\runningtitle{Deep Surveys}

\begin{document}


\section{Introduction}


Surveys with ISO (Kessler et al 1996), in particular with the CAM 
(Cesarsky et al 1996) and PHOT (Lemke et al 1996) instruments, will
greatly extend our understanding of extra-galactic populations and
their cosmological evolution.  The main advantages that ISO surveys
have over e.g IRAS are increased sensitivity/depth and wavelength
coverage.  Within the Guaranteed and Open Time programmes there are
many field surveys which will efficiently map the limits in these
parameters. In this talk I will briefly overview those surveys before
concentrating in more detail on one survey in particular, the ISO
survey of the Hubble Deep Field (HDF),
to illustrate the kind of results that can be expected.

\section{ISO Surveys}


Since ISO observes in the mid to far-infrared thermal dust emission
is the dominant process, although at higher redshifts and shorter
wavelengths star-light will play a role.

The extra-galactic populations that ISO will expect to see are
varied: normal galaxies at low-$z$; star-burst galaxies at
moderate $z$; ultra-luminous and hyper-luminous galaxies at high-$z$;
AGN selected in an unbiased fashion and forming
ellipticals or proto-spheroids at high-$z$. 
The ISO surveys will make significant advances in our
knowledge of all of these, although individual surveys will be better
suited to study some populations rather than others.


Surveys with ISO cover a full spectrum from very wide and shallow to
extremely narrow and deep, they also cover the full wavelength range
of ISO.  Table \ref{surveys} presents a reasonably complete list of
these surveys.  This table presents the area and integration time per
sky position.  Both of these figures are somewhat debatable for the
deeper surveys where multiple coverage gives different integration
times at different positions.  Integration time is used as an
indicator of depth, since not all of the surveys have yet published
sensitivity figures.

\begin{table}[htb]
\begin{center}
\caption{Field Surveys with ISO}\label{surveys}
\begin{tabular}{lllll}
\hline
Survey Name  & PI [refs.] & Wavelength & Integration & Area\\
             &    &   $/\mu$m  &   $/$s      & $/{\rm sq deg}$\\
\\
PHT Serendipity Survey & D. Lemke [1] & 175       & 0.5            & 7000 \\
CAM Parallel Mode & C. Cesarsky   [21]  & 7         & 150            & 33 \\
ELAIS         & M. Rowan-Robinson [16]  & 7, 15, 90 & 40, 40, 24     & 7, 12, 12\\
CAM Shallow   & C. Cesarsky       [4]  & 15        & 180            & 1.3  \\
IR Back       & K. Mattila            & 90, 135, 180 & 23, 27, 27  & 1, 1, 1 \\
FIR Back      & J-L. Puget            & 175       & 64             & 1  \\
SA 57         & H-U. N{\o}rgaard-Nielson [14] & 60, 90    & 150, 50        & 0.42  \\
CAM Deep      & C. Cesarsky        [4] & 7, 15, 90 & 800, 990, 144  & 0.28, 0.28, 0.28\\
Serendipitous comet fields       & C. Cesarsky           & 12        & 302           & 0.11\\
CAM Ultra-Deep& C. Cesarsky           & 7         & 3520           & 0.013 \\
ISOHDF South  & M. Rowan-Robinson     & 7, 15     &$>6400, >6400$  & 4.7e-3, 4.7e-3 \\
Deep SSA13    & Y. Taniguichi         & 7         & 34000          & 2.5e-3\\
Deep Lockman  & Y. Taniguichi     [22]  & 7, 90, 175& 44640, 48, 128 & 2.5e-3, 1.2 , 1 \\
ISOHDF North  & M. Rowan-Robinson [20]  & 7, 15     & 12800, 6400    & 1.4e-3, 4.2e-3 \\
\hline
\end{tabular}
\end{center}
\end{table}

\section{ISO HDF}

For a number of reasons the results of the deep surveys, in particular
the ISO HDF survey, were the first to be announced.  One reason
for this is technical; the multiple redundancy in the deep observations makes
them easier to interpret than the shallower surveys.  In the case of
the ISO HDF the pre-existing multi-wavelength data made detailed scientific
analysis possible without requiring ground-based follow-up programmes.
In addition, since this was an ISO Director's Time programme observing
an area of great international importance,
the proprietary period was shortened from 1 year to 3 months.
The first observations were performed in July 1996 and consisted of
three maps in both the 7 and 15 $\mu$m bands, centred on the 3 WFPC
centres of the HDF.  The 7 $\mu$m observations have recently
been repeated (July 1997) and  these observations will be
reported shortly.

The data reduction processes
have been described by Serjeant et al
(1997).
The main feature of the data reduction process is
that different pointings at the same sky position were median averaged
to filter out low level glitch effects.


The sources were extracted from the resulting maps using a standard
connected-pixel algorithm (Goldschmidt et al 1997).  Two lists were generated. the first, `complete', list was
generated fully automatically at higher SNR; while the second
`supplementary', list used a lower SNR threshold and sources rejected if they did not
look point-like on the ISO maps.  Reliability and completeness
estimates were made for the `complete' list using simulated data
and other techniques.


Some care had to be taken with the optical associations since the 
Airy disk of each ISO source covers many optical HDF galaxies.  Mann et al (1997) used a likelihood technique to
ensure highly reliable identifications.  This technique made no
assumptions about the optical magnitudes of the ISO sources,
merely that a positional coincidence with a bright optical source
is less likely to occur by chance (since they are rare) than
a coincidence with  a faint source (which are common).
The reliability of the associations was then assessed by comparison
with associations from a  catalogue of random positions.  
Any source that did not
have a reliable counterpart was excluded from further analysis. The
number of objects which failed this test in the `complete' lists
was consistent with the internal estimates from the source extraction
process.


Oliver {\em et al.} (1997a) used the  `complete'
source list to test two count models, that of Pearson \&
Rowan-Robinson (1996; PRR) and that of Franceschini {\em et al.} (1994).  
Both of these models fit the steep evolution seen in the IRAS 60 $\mu$m
catalogues.   The 15 $\mu$m data immediately confirmed that such
strongly evolving models were required (a no evolution model was ruled
out at 3 sigma).  The PRR model did not fit the 7 $\mu$m counts because
it predicted too many bright galaxies.  Both models comprised a
number of different populations and, with count data alone, it was
not possible to determine whether these populations were in the
correct proportions.


\begin{figure}
\begin{center}
\epsfig{file=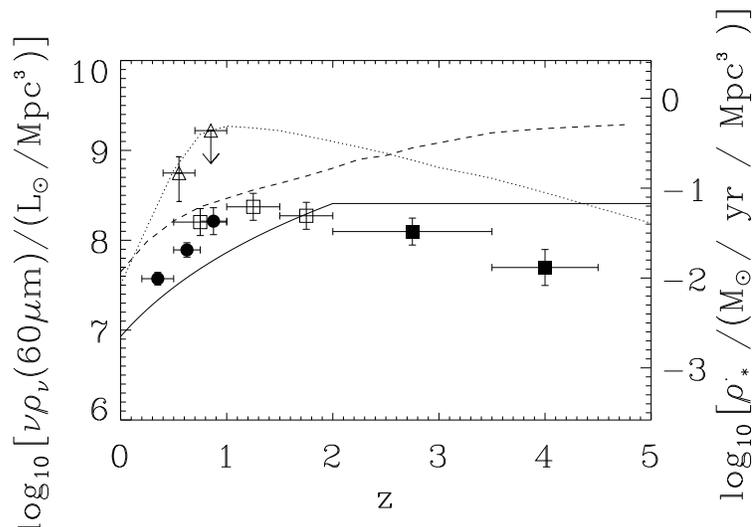,angle=90,width=9.5cm}
\caption{
The luminosity density at 60 $\mu$m and the star-formation rate 
as a function of redshift.  The solid
curve is derived from the PRR  model.  The broken curve is the model
of Franceschini et al (1994, 1997) and the dotted curve is one infall
model of Pei and Fall (1995).  The triangles are the estimates 
 derived directly from the ISO data (excluding
non-confirmed objects).  Also shown are
the star formation rates derived from the ultraviolet
luminosity density by Madau et al (1997) solid squares, Connolly et al
(1997) open squares and Lilly et al (1996) solid circles.}\label{sfr}
\end{center}
\end{figure}

The power of the optical HDF observations (Williams {\em et al.}
1996 and others) was to provide spectral
energy distributions, spectroscopic
redshifts and, where those were not available, photometric redshifts.  
This allowed Rowan-Robinson {\em et al.} (1997) to distinguish
star-forming galaxies with high mid-IR to optical ratios from
more normal objects.  The redshifts provided
luminosities, luminosity density and hence star-formation rate (Figure
\ref{sfr}). 
The star-formation rate determined from this analysis was considerably
higher than estimates from the optical/U-V (e.g. Madau {\em et al.}
1996) suggesting that dust obscuration plays a more important role
in star-forming regions than had been expected from the optical work.
The luminosity densities were also higher than the models
which were consistent with the 15 $\mu$m counts and the faint 7 $\mu$m
counts.

The discrepancy with the optical depends on the optical extinction
corrections applied. Recently Madau {\em et al.}  (1997) have applied
larger extinction corrections, but these are still not sufficient to
explain the discrepancy.  The apparent discrepancy with the count
models can have a number of explanations.  This analysis used mainly 7
$\mu$m sources and included objects from the `supplementary' list, it
is possible that these fainter sources tell a different
story. Alternatively the extrapolation from short to long wavelengths
may be at fault.  More likely is that the addition of the optical
data, breaks the degeneracy between populations in the source count
models.  (A recent re-analysis of the ISO data and the new 1997 7
$\mu$m data has found that two of the sources used in this
calculation are spurious and if these objects are removed the
significance of the result is reduced somewhat, nevertheless new
objects are also discovered, and a full discussion of these new
results will appear shortly.)  These results are very exciting for our
understanding of the history of star-formation in the Universe, it
will be very interesting to see if they are confirmed in more detail
by the other ISO surveys.  It should be noted the importance of the
optical work in establishing this result demonstrating that optical
follow-up will be of crucial importance to the other ISO surveys.

Many more details can be found on `The ISO-HDF Project WWW Site'
(Mann R.G., Oliver S.J., Serjeant S.B.G., Goldschmidt P., Gruppioni
C., 1997; http://athena.ph.ic.ac.uk/hdf/)
including:  active source maps; a guided tour through
the data reduction steps; and some new tests of the data reduction 
procedures.

\end{document}